\documentclass[12pt,fleqn]{article}
\usepackage{exscale}
\usepackage{epsfig}

\newcommand{\kf}{{\bf k}}
\newcommand{\Rf}{{\bf R}}

\newcommand{\Asf}{\mathrm{A}}
\newcommand{\asf}{\mathrm{a}}
\newcommand{\Bsf}{\mathrm{B}}
\newcommand{\bsf}{\mathrm{b}}

\newcommand{\fsf}{{\mathrm f}}

\newcommand{\Psf}{{\mathrm P}}
\newcommand{\Qsf}{{\mathrm Q}}

\begin{document}


\title{Perturbation expansion for 2-D Hubbard model}

\author{V. Zlati\'{c}, B. Horvati\'{c}
and B. Doli\v{c}ki \\
{\em Institute of Physics, Bijeni\v{c}ka cesta 46,
P. O. Box 304,} \\
{\em HR--10001 Zagreb, Croatia}
\and
S. Grabowski and P. Entel \\
{\em Gerhard-Mercator University, D--47048 Duisburg,
Germany}
\and
K.--D. Schotte \\
{\em Institut f\"{u}r Theoretische Physik, Freie
Universit\"{a}t Berlin,} \\
{\em  Arnimallee 14, D--14195Berlin, Germany}}

%
%
\maketitle


\begin{abstract}
We develop an efficient method to calculate the
third-order corrections to the self-energy of
the hole-doped two-dimensional Hubbard model in
space-time representation. Using the Dyson equation
we evaluate the renormalized spectral function
in various parts of the Brillouin zone and find
significant modifications with respect to the
second-order theory even for rather small values
of the coupling constant $ U $. The spectral function
becomes unphysical for $ U \simeq W $, where $ W $
is the half-width of the conduction band. Close to
the Fermi surface and for $ U<W $, the single-particle
spectral weight is reduced in a finite energy
interval around the Fermi energy. The increase of
$ U $ opens a gap between the occupied and unoccupied
parts of the spectral function.

\end{abstract}

\section*{Introduction}
The hole-doped two-dimensional (2-D) Hubbard model
continues to attract considerable attention as it
is believed to capture the essential features
of ``underdoped cuprates''~\cite{anderson.97}.
In particular, the spectral properties of the model
at small doping and the coupling strength $ U\geq 2W $,
where $W$ is the half-width of the conduction band,
are thought to be relevant for the angular
resolved photoemission spectroscopy
({\sc arpes}) experiments.
The {\sc arpes} data are used to study the
frequency and momentum dependence of the
single-particle excitations of 2-D electrons in
$ CuO_2 $ layers~\cite{valla.99,kaminski.00}
but the interpretation
of the experimental results is quite difficult
and often controversial.
On the one hand, the observed spectral features
do not fit any simple conceptual framework;
on the other hand, reliable theoretical results
are not at hand. Thus, it is of interest to study
in some detail the effect of correlation on the
spectral function of the hole-doped 2-D Hubbard
model.

The weak-coupling analysis of renormalized
single-particle excitations has been presented
in a number of papers
\cite{virosztek.90}--\cite{yamada.99},
which treat $ U $ as an expansion parameter and
consider the effects of correlation, doping, and
temperature in various parts of the Brillouin zone.
These papers show that correlation changes
significantly the single-particle spectral properties
even for relatively small values of $ U $, and the results
exhibit a number of interesting features which are also
seen in cuprates. However, a quantitative comparison
with the experimental data is difficult
since most weak-coupling theories become
unreliable for $ U \geq W $.

The breakdown of the weak-coupling schemes
based on truncated perturbation expansions is not
immediately discernible from the spectral function,
but is signalled for $U\simeq W$ by the negative
compressibility \cite{zlatic.97} and the rapid
deviation of the Fermi volume $ v_F $ from the values
required by the Luttinger theorem \cite{halbot.97}.
The theories based on an infinite summation of
selected classes of diagrams
\cite{fukuyama.91,randeria.92,langer.95}
are also unreliable for large values of $ U $,
because they overemphasize the spectral weight of
quasiparticle peaks and do not produce the
Hubbard side-bands, which are typical of strong local
correlations. Thus, the perturbational results
obtained so far give some insights
into the properties of the Hubbard model
but do not allow a consistent description of
correlated electrons at intermediate or large
values of $ U $ which one needs to discuss the
cuprates or make a comparison with the t-J model.

The weak-coupling approach to the Hubbard model
poses a number of questions which should be
considered if the results are to be extrapolated
into the large-$ U $ regime.
What is the range of validity of the asymptotic
$ U $ expansion?
What is the importance of the terms that are
neglected in the self-energy expansion?
Is it possible to use a finite-order perturbation
theory for the values of $ U $ such that the
low-energy excitations of the Hubbard model and
the t-J model look similar?
Are the weak-coupling results obtained by perturbative
methods representative of the strong-coupling limit?
And finally, does the low-energy physics of the
2-D Hubbard model, as defined by some weak-coupling
scheme, produce the right phenomenology for underdoped
cuprates?

To answer these questions one would have to examine the
general structure of the perturbation expansion or
compare the weak-coupling solution with the exact one,
which is not possible at present.
Some insight, however, can be obtained by extending
the perturbation expansion beyond the 2nd order and
studying the stability of a truncated series with
respect to higher-order corrections.
Here, we calculate the momentum-dependent
single-particle self-energy up to the 3rd order,
and show that the individual 3rd-order diagrams
are large but that the total 3rd-order contribution
is much smaller than the 2nd-order one.
This approximate mutual cancellation
of 3rd-order diagrams holds for any doping and not
just for zero doping, where it is a consequence
of electron-hole symmetry.

Once the 3rd-order contribution becomes comparable
to the 2nd-order one, which happens here for
$ U \simeq W $, the truncated perturbation series
leads to unphysical results. In the ``physical'' range,
$ U < W $, we find a number of interesting features
which offer additional insight into the anomalies
of the 2-D hole-doped Hubbard model.
However, the error of neglecting the higher-order
terms might become significant for larger values of
$ U $ even within the physical range \cite{comment}.
It would be interesting to see whether the 4th order
contribution improves the perturbation theory and 
extends it to experimentally relevant values of $U$,
or whether it renders the weak-coupling approach
useless. The technical problems involved with
such calculations are not much different from the
problems encountered in the 3rd-order calculations
\cite{horvatic.2001}.
The Matsubara summations for the 4th-order diagrams
are straightforward, if tedious,
and the 4th-order momentum summations follow from
the same numerical strategies which solve the 3rd
order. Thus, the 3rd-order calculations can be
considered as a small but necessary step in our
efforts to clarify the properties of the Hubbard model
up to intermediate values of the coupling strength.

We should also remark that the accurate
characterization of the weak-coupling regime might
be of some interest for the approximate schemes
which interpolate between the small-$ U $ and the
large-$ U $ limit of the model. In the case of the
infinite-dimensional Hubbard model and the
single-impurity Anderson model, interpolations
like that \cite{martin-rodero.82}--\cite{kotliar.96}
come very close to the exact solution.
However, on a 2-D lattice one deals with the
anisotropic {\bf k} space, and the interpolation
schemes might be difficult to construct.

The present paper is organized as follows.
First the retarded 2nd- and 3rd-order self-energies
are expressed in terms of multiple momentum integrals.
Then we introduce the space-time representation and use
the Fast Fourier Transform ({\sc fft}) algorithm to
evaluate the self-energy as a function of energy for
all points in the Brillouin zone. The relative
importance of the 2nd- and 3rd-order terms obtained
by {\sc fft} is analyzed and the stability of the
weak-coupling solution is discussed.
Next, the spectral properties of the model are calculated
for low temperatures, for a given value of the chemical
potential, and for various points in the Brillouin zone.
Finally, the spectral features and their relevance for
the experimental data are briefly discussed.
The calculations are explained in detail in the
Appendix.

\section*{Calculations}

To start with, the Hubbard Hamiltonian is written as
\begin{equation}
H = \sum_{i,j, \sigma} t_{ij}
c^{\dagger }_{i \sigma} c_{j \sigma}
\, - \,\mu' \sum_{i, \sigma} n_{i \sigma}
\, + \, U \sum_{i=1}^{N_{g}}
(n_{i \uparrow }
- <\!\! <\! n_{i \uparrow} \!>\!\! >)
(n_{i \downarrow }
- <\!\! <\! n_{i \downarrow } \!>\!\! >)\;\;\; ,
                 \label{eq:H_Hubbard}
\end{equation}
where $t_{ij}$ is the nearest-neighbor hopping
integral, $c_{i\sigma}$ $(c^{\dagger }_{i\sigma})$
destroys (creates) an electron at site $ \Rf_{i} $
with spin $ \sigma $,
$n_{i\sigma}=c^{\dagger }_{i\sigma}c_{i\sigma}$ is
the electron number operator, $ U $ is the local
electron-electron interaction, and
$ <\!\! <\! \cdots \!>\!\! > $ denotes the
ensemble average over the eigenstates of the
{\em full\/} Hamiltonian (\ref{eq:H_Hubbard}).
The parameter
$ \mu'= \mu \, - <\!\! <\! n_{- \sigma } \!>\!\! >U $
is the ``effective chemical potential'', $ \mu $
being the chemical potential proper. The energy of
an unrenormalized single-particle excitation
propagating with wave vector $ \kf $, counted from the
Fermi level, is
$ \omega_{\kf}^{0} = \epsilon_{\kf} - \mu' $,
where
$\epsilon_{\kf} = - 2 [t_{x} \cos(k_{x} a_{x})
+ t_{y} \cos(k_{y} a_{y})] $.
The Fermi momentum is denoted by $ \kf_{F} $.
We consider the fluctuations above a
mean-field-like paramagnetic state, in which the
number of particles coincides with the exact
particle number,
$ <\!\! <\! n_{i\uparrow} \!>\!\! > \, =
\, <\!\! <\! n_{i\downarrow } \!>\!\! >
\, = n_{e}/2 $,
and assume $ t_x = t_y = $ t, the half-bandwidth
being $ W = 4$t.

The self-energy diagrams are generated by the usual
Matsubara imaginary-time perturbation expansion with
respect to the last term in Eq.\ (\ref{eq:H_Hubbard}).
All the 2nd- and 3rd-order diagrams are shown in
Fig.\ 1, where the dashed line represents the local
interaction ($ -U $) and the full line stands for the
unperturbed propagator
$ G^{0}_{\kf}(i\omega_{n})
= (i\omega_{n} - \omega^{0}_{\kf})^{-1} $,
defined with the first two terms in
Eq.\ (\ref{eq:H_Hubbard}). Note that
there are no self-energy diagrams
that are {\bf k} and $ \omega $ independent,
the so-called one-legged diagrams. Because of the
separation of the Hamiltonian as in
(\ref{eq:H_Hubbard}), all one-legged diagrams of all
orders add up to zero. Equivalently one could say
that they have been hidden in the ``effective chemical
potential''
$ \mu'= \mu \, - <\!\! <\! n_{- \sigma } \!>\!\! >U $.

The self-energy calculations for a finite lattice
with periodic boundary conditions and a discretized
time axis simplify considerably in the space-time
representation \cite{schweitzer.91}. The 2nd- and
the 3rd-order {\em retarded\/} proper self-energy
contributions are given by the expressions
\begin{eqnarray}
{\Sigma}^{(2)}_{\Rf}(t)
& = & -i \, U^{2} \, \Theta(t)
\, \left[ \asf^{2}(\Rf , t)
\, \bsf^{\ast}(\Rf , t)
+ \bsf^{2}(\Rf , t)
\, \asf^{\ast}(\Rf , t) \right] \;\;\; , \\
{\Sigma}^{(3)}_{\Rf}(t)
& = & \, U^{3} \, \Theta(t)
\, \left\{ \left[ \asf(\Rf , t)
\, {\mathrm w}_{1}(\Rf , t)
- \bsf(\Rf , t)
\, {\mathrm w}_{1}^{\ast}(\Rf , t)
\right] \right. \nonumber \\
& & \makebox[1.2cm]{}
- \left. \left[ \asf^{\ast}(\Rf , t)
\, {\mathrm w}_{2}(\Rf , t)
+ \bsf^{\ast}(\Rf , t)
\, {\mathrm w}_{3}(\Rf , t)
\right] \right\} \;\; ,
\end{eqnarray}
where $ \asf(\Rf , t) $ and  $ \bsf(\Rf , t) $ are
the standard \cite{kadanoff.61} double-time Green's
functions of $ H_{0} $,
\begin{eqnarray}
\asf(\Rf , t)
& \equiv & <\! c_{\bf 0}^{\dagger }(0)
\, c_{\Rf}(t) \! >
\; = \frac{1}{N} \sum_{\kf}
e^{i(\kf \cdot \Rf- \omega^{0}_{\kf} t)}
\, \fsf(\omega^{0}_{\kf}) \;\; ,\\
\bsf(\Rf , t)
& \equiv & <\! c_{\Rf}(t)
\, c_{\bf 0}^{\dagger }(0) \! >
\; = \frac{1}{N} \sum_{\kf}
e^{i(\kf \cdot \Rf- \omega^{0}_{\kf} t)}
\, \fsf(- \omega^{0}_{\kf}) \;\; ,
\end{eqnarray}
while the functions $ {\mathrm w}_{i}(\Rf , t) $
are space-time convolutions of functions composed
of products of $ \asf(\Rf , t) $'s and
$ \bsf(\Rf , t) $'s,
given by expression (\ref{eq:A7}) in the Appendix.
Instead of evaluating these convolutions directly,
we decouple them by a pair of Fourier transforms,
as shown in (\ref{eq:A8}). (The time variable $ t $
should not be confused with the hopping amplitude t.)

The {\em retarded\/} self-energy in the
energy-momentum representation,
$ \Sigma_{\kf}(\omega )
\equiv \Sigma_{\kf}(\omega + i 0^{+}) $,
is then given by the inverse Fourier transform
\begin{equation}
\Sigma_{\kf}(\omega )
= \sum_{\Rf } e^{-i \kf \cdot \Rf}
\int_{0}^{\infty } dt \; e^{i\omega t}
\left[ \Sigma _{\Rf}^{(2)}(t)
+ \Sigma_{\Rf}^{(3)}(t) \right] \;\; .
\label{eq:sigma_t}
\end{equation}
The self-energy calculation is thus reduced to
a sequence of Fourier transforms, which can be
evaluated very efficiently by the {\sc fft}
technique.
In this paper we consider a lattice with
$ N_{g} = 256 \times 256 $ sites and define
discretized momenta in the quadratic Brillouin
zone as
$ \kf = (k_{x}, k_{y}) $, where
$ k_{x,y} = \Delta k(l_{x,y}-1) $ with
$ \Delta k = {2\pi}/{\sqrt{N_{g}}} $ and
$ l_{x,y} = 1,\ldots ,\sqrt{N_{g}} $.
The $ \Gamma $ point is at $ \kf = (0,0) $,
the X point at $ \kf = (\pi , 0) $,
the M point at $ \kf = (\pi/2 , \pi/2) $, and
the Z point at $ \kf = (\pi , \pi) $.

Having found $ \Sigma_{\kf}(\omega,T) $, we calculate
the spectral function $ A_{\kf}(\omega ,T)$ from the
Dyson equation,
\begin{equation}
A_{\kf}(\omega , T)
= - \frac{1}{\pi} \,\rm{Im}
\; \frac{1}{\omega +i0^{+} - \omega_{\kf}^{0}
- \Sigma_{\kf}(\omega , T)} \;\;\; ,
                    \label{eq:A_spectrum}
\end{equation}
and obtain the renormalized density of states,
\begin{equation}
\rho (\omega , T)
= \frac{1}{N} \sum_{\kf} A_{\kf}(\omega,T),
                       \label{eq:rho}
\end{equation}
and the renormalized particle number,
\begin{equation}
n_{e} = 2 \int_{-\infty}^{\infty}
d\omega \, \fsf(\omega)
\, \rho(\omega , T) \;\;\; .
                       \label{eq:nn}
\end{equation}

Eq.\ (\ref{eq:nn}), together with Eqs.\ (\ref{eq:rho})
and (\ref{eq:A_spectrum}), establishes the functional
dependence of the renormalized particle number
$ n_{e} $ on the ``effective chemical potential''
$ \mu' $, with $ U $ and $ T $ as parameters,
everything scaled in units of $ W $:
\begin{equation}
n_{e} = {\cal N}(\mu'/W ; U/W, k_{B}T/W)
\label{eq:nN}
\end{equation}
On the other hand, the pure chemical potential $ \mu $
is given by
\begin{eqnarray}
\mu & = & \mu' + \frac{U}{2} n_{e} \nonumber \\
    & = & \mu' + \frac{U}{2}
{\cal N}(\mu'/W ; U/W, k_{B}T/W)
\label{eq:mu}
\end{eqnarray}
Eqs.\ (\ref{eq:nN}) and (\ref{eq:mu}) establish the
dependence of $ n_{e} $ on $ \mu $ or vice versa,
with both quantities given parametrically as
functions of $ \mu' $. In this way one does not have
to really {\em solve\/} Eq.\ (\ref{eq:nn}) as a
self-consistency {\em equation\/} for $ n_{e}(\mu ) $,
it suffices to just {\em evaluate\/}
Eqs.\ (\ref{eq:nN}) and (\ref{eq:mu}) for a number
of close-lying values of $ \mu' $.

One should note that the $n$-consistency is here
forced upon the approximate
proper self-energy,
$ \Sigma_{\kf}^{(2)}(\omega)
+ \Sigma_{\kf}^{(3)}(\omega) $,
and can therefore be attained only approximately.

The peaks of $ A_{\kf}(\omega) $ give
the dispersion of quasiparticle excitations,
$ \omega_{\kf} $; the renormalized Fermi surface
is defined by the set of points in the
momentum space at which $ \omega_{\kf} = 0 $.
(At the Fermi surface, $ A_{\kf}(\omega) $ has a
singularity at the Fermi energy $E_F$.)
According to the Luttinger theorem,
the number of $ \kf $ points enclosed by the
Fermi surface (i.e. the Fermi volume $ v_{F} $)
should coincide with $ n_{e} $. In our approximate
treatment, which is based on the 3rd-order self-energy
and the Dyson equation,
we find $ v_{F} > n_{e} $, but the relative
difference between $ v_{F}(U, \mu) $ and
$ n_{e}(U, \mu) $ for $ U \leq W $ is very small.

In what follows, we first discuss
$ \Sigma_{\kf}(\omega , T) $ and then the ensuing
spectral function, the density of states,
the renormalized dispersion, and the Fermi surface
of the model for the temperature $ k_{B}T = $ t/250,
$ U = 3.5 $t, and $ \mu' = 0.2 $t,
which corresponds to $n_e=0.96$. The same
numerical program, available from the authors
upon request, returns the self-energy in the
irreducible wedge of the Brillouin zone for
any other value of $T$, $U$, and $\mu$.

\section*{Results and discussion}

In Figs.\ 2 and 3 we show the real and the imaginary
part of $ \Sigma^{(3)}_{\kf}(\omega) $, respectively,
plotted versus $\omega$ for several momenta along the
$ \Gamma $--X--Z--M--$ \Gamma $ cuts through the
Brillouin zone. The dotted and the dashed-dotted lines
show the two 3rd-order contributions,
$ \Sigma^{(3)}_{\mathrm{PH}}(\omega) $ and
$ \Sigma^{(3)}_{\mathrm{PP}}(\omega) $, respectively,
while the full line is their sum,
$ \Sigma^{(3)}_{\kf}(\omega) $.
The magnitudes of the individual 3rd-order
contributions are about the same as the magnitude of
$\Sigma^{(2)}_{\kf}(\omega)$
(see below and in Ref.\ \cite{zlatic.95}) but
the total 3rd-order contribution is relatively
small, except in a narrow frequency range where the
approximate cancellation of electron-electron
and electron-hole terms does not occur.
This behavior indicates that the correct solution of
the hole-doped model requires all 3rd-order diagrams,
and can not be obtained by the partial summation of
electron-electron or electron-hole diagrams.

The functional form of $\Sigma_{\kf}^{(3)}(\omega) $
is very much {\bf k}-dependent.
Far off the Fermi surface
$ {\mathrm{Im}}\Sigma_{\kf}^{(3)}(\omega) $
does not show much structure in the low-energy region
and its slope around $ E_{F} $ is always small
(see Figs.\ 3 and 5 for data corresponding to
$ \Gamma $ and Z points).
On the other hand, for $ \kf \simeq \kf_F $,
${\mathrm{Im}}\Sigma_{\kf}^{(3)}(\omega) $
acquires a pronounced minimum at $E_F$
(see Figs.\ 3 and 5 for data corresponding to
X and M points). In the low-energy region,
the $ {\mathrm{Im}}\Sigma_{\kf}^{(3)}(\omega)$
and $ {\mathrm{Im}}\Sigma_{\kf}^{(2)}(\omega)$
have opposite sign, and both vanish at $E_F$
\cite{error} with a zero slope.

The properties of the ``full'' self-energy,
$ \Sigma_{\kf}^{(2)}(\omega)
+ \Sigma_{\kf}^{(3)}(\omega) $,
are summarized in Fig.\ 4, where
$ {\mathrm{Im}}\Sigma_{\bf k}(\omega) $
is plotted versus $\omega$ for the same {\bf k}'s
as in Fig.\ 3. The low-energy behavior
at $\Gamma$, X, Z, and M is displayed in Fig.\ 5.
For comparison, the individual 2nd- and 3rd-order
contributions are shown as well.
The low-energy features of the
${\mathrm{Re}}\Sigma_{\bf k} (\omega)$
are not changed much by the 3rd-order
renormalization, regardless of ${\kf}$.
On the other hand, the low-energy properties of
${\mathrm{Im}}\Sigma_{\bf k} (\omega)$
are strongly
influenced by the 3rd-order corrections.
For example, close to X and M points, the 2nd- and
the 3rd-order terms nearly cancel and make the
low-energy part of
$ {\mathrm{Im}}\Sigma_{\bf k}(\omega) $
rather flat in an extended interval around $E_F$.
The relative contribution of the 2nd- and the
3rd-order terms depends on the coupling strength and
for a given $\mu$ there is always
a critical correlation $U_{c}(\mu,{\kf}_c)$ such
that the coefficient of the $ \omega^{2} $ term in
$ {\mathrm{Im}}\Sigma_{{\kf}_c}(\omega) $ vanishes
at some ${\kf}_c$.
For $ U < U_c$ the curvature of
$ {\mathrm{Im}}\Sigma_{\kf}(\omega) $ at $E_F$ is
negative for all ${\kf}$
and the ensuing spectral weight is
always positive. However, for $ U > U_c $ the curvature
of $ {\mathrm{Im}}\Sigma_{\bf k} (\omega) $ at $E_F$
becomes positive at some points in the Brillouin zone,
and the corresponding spectral weight becomes negative.
This behavior indicates that
the 3rd-order $ U $ expansion breaks down for $U\simeq W$,
and that $\Sigma^{(2)}_{\kf}(\omega)$ provides
an accurate renormalization for small values of $U$  only.

For $U$ close to but less than $U_c$, the curvature
of ${\mathrm{Im}}\Sigma_{\kf}(\omega)$ at $E_F$ is
found to be the smallest for ${\kf}\simeq {\kf}_F$.
Thus, in the presence of correlations the Fermi surface
properties assume qualitatively new features due to the
$\omega^3$ and higher-order self-energy terms.
Unfortunately,
these non-Fermi-liquid ({\sc nfl}) features require
large values of $U$ and can not be properly discussed
without the higher-order terms in the expansion.
Leaving the 4th-order renormalization for future
studies \cite{horvatic.2001}, we consider here the
3rd-order renormalization and discuss spectral
properties for $ U < U_c \simeq W $.

The variation of $ A_{{\bf k}}(\omega )$ along
$ \Gamma \rightarrow {\rm X} $,
$ {\rm X} \rightarrow {\rm Z} $ and
$ {\rm Z} \rightarrow \Gamma $ cuts in the Brillouin
zone is shown in Fig.\ 6. The 3rd-order spectral
function, like the 2nd-order one \cite{zlatic.95},
assumes at all wave vectors a typical shape with a
low-energy quasiparticle peak and a high-energy
incoherent background.
The correlation reduces the quasiparticle spectral
weight and enhances the incoherent background of
$ A_{\kf}(\omega) $.
However, for $ U = 3.5 $t and $ n_{e} = 0.96 $
the transfer of spectral weight out of the
low-energy region is small, and the quasiparticle peak
is not fully separated from the incoherent background,
except for $\kf$'s which are far off the Fermi surface.
At the Fermi surface, the
singular quasiparticle peak can be represented as
$ A_{{\bf k}_F}(\omega )
= Z_{{\kf}_F}\,\delta (\omega - E_{F})$,
where $ Z_{\bf k} $ describes the reduction of the
quasiparticle weight due to self-energy corrections.
For $ \kf \neq \kf_{F} $ the quasiparticle peak
broadens, becomes asymmetric, and attains a maximum
at some finite frequency.
The asymmetry of $ A_{\kf}(\omega)$ is caused by an
incoherent background, which slows down the
high-$\omega$ decay, and the {\sc nfl} behavior at
small $\omega$.
A new feature due to the 3rd-order, which begins to emerge
for $U\simeq W$, is the suppression of spectral weight
in a finite energy interval around $ E_{F} $.
For ${\kf}\simeq {\kf}_F$, the singular quasiparticle
peak at the Fermi energy is separated from the occupied
and unoccupied incoherent states by a pseudogap.
For $ \kf < \kf_{F} $ a real gap appears
between the (occupied) quasiparticle peak
and the (unoccupied) incoherent states.
(The small background seen in the numerical data is
due to the damping factor used for the real-axis
propagators.)

The renormalized quasiparticle dispersion,
$ \omega_{\kf} $, is obtained by tracing the maximum of
$ A_{\kf}(\omega) $ across several cuts in the momentum
space and is shown in Fig.\ 7. Circles show
$ \omega_{\kf} $ and full line gives the unrenormalized
dispersion, $ \omega_{\kf}^{0} $, for the same number
of particles. Correlation reduces the overall width
of the quasiparticle band and extends the flat
dispersion around $ (\pi,0) $. Essentially the same
dispersion is obtained by solving the secular equation
\begin{equation}
\omega_{\kf}(T) -(\epsilon_{\kf} - \mu')
- \Sigma_{\kf}[\omega_{\kf}(T), T] = 0 \;\; .
             \label{eq:secular}
\end{equation}
The inspection of Figs.\ 6 or 7 shows that
the M-point quasiparticle peak is somewhat above
$ E_{F} $, while the X-point peak is right at
$ E_{F} $. Thus, the renormalized Fermi surface,
defined by the set of {\bf k} points such that
the quasiparticle peak is at $E_{F}$,
has a different topology than the non-interacting
Fermi surface with the same number of holes.
The renormalized Fermi surface resembles the
tight-binding result which adds to $ t_{ij} $
the next-nearest-neighbor hopping $t_{ij}^{'}$.

The renormalized density of states $\rho(\omega)$
calculated for $ U = 3.5$t and $ n_{e} = 0.96 $
is plotted in Fig.\ 8 as a function of $\omega$.
The transfer of the low-energy spectral weight
out of the low-$ \omega $ region is clearly seen
but $U\leq W$ is not sufficient to generate the
incoherent Hubbard side-bands.
The density of states at $ E_{F} $ is not much
enhanced from the unperturbed value despite the
reduced dispersion, because the quasiparticle weight
is reduced by $ Z_{\bf k} $. One has a metal but
a strange one.

In summary, we developed an efficient method to
perform the momentum summations in the self-energy
expansion for the hole-doped 2-D Hubbard model and
evaluated the coefficients of the $U^2$ and the $U^3$ terms.
The third-order corrections lead to additional
anisotropy and asymmetry of the spectral function
and give rise to new features with respect
to the 2nd-order renormalization. For large enough
$U$ we find strong reduction of the single-particle
spectral weight around $E_F$.
The quasiparticle dispersion is also reduced and the
saddle point singularities are extended.
The Fermi surface obtained by the 3rd-order renormalization
for a hole-doped system is closed around the Z point.
At the Fermi surface, the quasiparticle peak
and the incoherent background are separated
by a pseudogap. Off the Fermi surface we find
a finite region of negligible spectral weight
between the quasiparticle peak and the Fermi energy.
This behavior hints to a possible scenario in which
the system is metallic with a narrow Kondo-like resonance
at the Fermi energy, and a pseudogap for incoherent
excitations.
However, the anomalous spectral features which we obtain
here are due to the cancellation of the 2nd- and the
3rd-order self-energy terms, and the stability of
this result with respect to the 4th-order
renormalization remains to be seen.
The 3rd-order perturbation expansion shows that
the solution of the hole-doped model
can not be obtained by partial summation of diagrams,
and that the properties inferred from the 2nd-order
theory seem to be qualitatively correct for $ U\leq W $
only.

\section*{Appendix}

Carrying out the Matsubara summations over Fermi
frequencies and continuing analytically the result
from the discrete points on the imaginary axis onto
the complex plane, we find

\begin{equation}
\Sigma_{\kf}^{(2)}(z)
= \left( \frac{U}{N} \right)^2
\sum_{\kf_{1}, \kf_{2}}
\frac{\Psf(\omega^{0}_{\kf - \kf_1},
\omega^{0}_{\kf_1 - \kf_2},
 \omega^{0}_{\kf_2})}
{z - ( \omega^{0}_{\kf - \kf_1}
+ \omega^{0}_{\kf_1 - \kf_2}
- \omega^{0}_{\kf_2})} \;\; .
\label{eq:2nd}
\end{equation}
The third-order self-energy
$ \Sigma_{\kf}^{(3)}(z) $ comprises two diagrams,
a particle-hole one and a particle-particle one,
\begin{equation}
\Sigma_{\kf}^{(3)}(z)
= \Sigma^{(3)}_{\mathrm{PH}}(\kf,z)
+ \Sigma^{(3)}_{\mathrm{PP}}(\kf,z) \;\; ,
\label{eq:3rd}
\end{equation}
where
\begin{eqnarray}
\Sigma^{(3)}_{\mathrm{PH}}(\kf,z)
& = & 2\left ( \frac{U}{N} \right)^3
\sum_{\kf_{1}, \kf_{2}, \kf_{3}}
\frac{\Psf( \omega^{0}_{\kf - \kf_1},
- \omega^{0}_{\kf_1 - \kf_2},
\omega^{0}_{\kf_2})}
{z - ( \omega^{0}_{\kf - \kf_1}
- \omega^{0}_{\kf_1 - \kf_2}
+ \omega^{0}_{\kf_2})}
\nonumber \\
& & \makebox[2.3cm]{}
\times \frac{\Qsf(- \omega^{0}_{\kf_1 - \kf_3},
\omega^{0}_{\kf_3})}
{(\omega^{0}_{\kf_1 - \kf_3} - \omega^{0}_{\kf_3})
- (\omega^{0}_{\kf_1 - \kf_2}
- \omega^{0}_{\kf_2})} \;\; ,
\label{eq:PH}
\end{eqnarray}
while the analogous expression for
$ \Sigma^{(3)}_{\mathrm{PP}}(\kf,z) $ is obtained
from $ - \Sigma^{(3)}_{\mathrm{PH}}(\kf,z) $ by
changing the sign of three $ \omega^{0}_{\kf} $'s
out of five,
viz.\ $ \omega^{0}_{\kf - \kf_1} $,
$ \omega^{0}_{\kf_1 - \kf_2} $, and
$ \omega^{0}_{\kf_1 - \kf_3} $,
everything else being the same as in (\ref{eq:PH}).
The functions $\Psf$ and $\Qsf$ are defined as
\begin{eqnarray*}
\Psf(\epsilon_1,\epsilon_2,\epsilon_3)
& = & \fsf(\epsilon_1)\,\fsf(\epsilon_2)
\,\fsf(\epsilon_3)
+ \fsf(-\epsilon_1)\,\fsf(-\epsilon_2)
\,\fsf(-\epsilon_3) \;\; ,\\
\Qsf(\epsilon_1,\epsilon_2)
& = & \fsf(\epsilon_1)\,\fsf(\epsilon_2)
- \fsf(-\epsilon_1)\,\fsf(-\epsilon_2) \;\; ,
\end{eqnarray*}
$ \fsf(\epsilon) = (e^{\beta \epsilon}
+ 1)^{-1} $ is the Fermi function, with
$ \beta = 1/k_{B}T $, and
$ \omega^{0}_{\kf} = \epsilon_{\kf} - \mu' $.

The above equations are not suitable for numerical
evaluation of
$ \Sigma_{\kf}^{(2)}
(\omega \pm i 0^{+}) $
and $ \Sigma_{\kf}^{(3)}
(\omega \pm i 0^{+}) $ because of two
reasons. First, one would have to deal with
four-dimensional and six-dimensional integrations in
{\bf k} space, respectively. Also, the denominators
in expressions (\ref{eq:2nd}) and (\ref{eq:PH})
vanish along various closed contours in {\bf k} space.
The multiple integrals are therefore defined by their
principal values and their evaluation requires a dense
grid in {\bf k} space, which makes the numerical
procedures very time-consuming.

The second- and third-order self-energies, given by
Eqs.\ (\ref{eq:2nd}) -- (\ref{eq:PH}), are therefore
transformed here from the energy-momentum to the
space-time representation, which allows an accurate and
efficient evaluation by the {\sc fft} algorithm.

In order to establish transparent shorthand notation,
we first define the Fourier transforms as the
operators
\begin{eqnarray*}
{\cal F}_{\kf \rightarrow \Rf} [ \cdots ]
& \equiv & \frac{1}{N} \sum_{\bf k}
e^{i \kf \cdot \Rf} \cdots \;\; , \\
{\cal F}_{\omega \rightarrow t} [ \cdots ]
& \equiv & \int_{-\infty}^{\infty}
\frac{d\omega }{2\pi}
\; e^{-i \omega t} \cdots \;\; ,
\end{eqnarray*}
as well as the inverse transforms
\begin{eqnarray*}
{\cal F}^{-1}_{\kf \rightarrow \Rf} [ \cdots ]
& \equiv & \sum_{\bf R}
e^{- i \kf \cdot \Rf} \cdots \;\; , \\
{\cal F}^{-1}_{\omega \rightarrow t} [ \cdots ]
& \equiv & \int_{-\infty}^{\infty}
dt \; e^{i \omega t} \cdots \;\; .
\end{eqnarray*}
We also introduce shorthand for four-dimensional
transforms
\[
{\cal F} \equiv {\cal F}_{\kf \rightarrow \Rf}
\: {\cal F}_{\omega \rightarrow t}
\makebox[1.5cm]{and}
{\cal F}^{-1} \equiv {\cal F}^{-1}_{\kf \rightarrow \Rf}
\: {\cal F}^{-1}_{\omega \rightarrow t} \;\; .
\]
The retarded/advanced self-energy in space-time
representation is then defined as
\[
\Sigma_{r/a}(\Rf, t) = {\cal F}
\left[ \Sigma_{\kf}
(\omega \pm i 0^{+}) \right] \;\; .
\]

For the 2nd-order self-energy we start from the
expression (\ref{eq:2nd}) for
$ \Sigma_{\kf}^{(2)}
(\omega \pm i 0^{+}) $
and first evaluate the $ \omega \rightarrow t $
Fourier transform. Using the relation
\[
{\cal F}_{\omega \rightarrow t}
\left[ \frac{1}{\omega \pm i \delta - \epsilon}
\right] = \mp i \, \Theta(\pm t) \, e^{- i
(\epsilon \mp i \delta ) t} \;\; ,
\]
which holds for any $ \delta > 0 $, we factorize
the denominator in (\ref{eq:2nd}) into a product
of three exponentials and find
\begin{equation}
\Sigma^{(2)}_{r/a}(\Rf, t)
= \mp i \, U^{2} \, \Theta(\pm t)
\, {\cal F}_{\kf \rightarrow \Rf}
\left[ S^{(2)}_{\kf} (t) \right] \;\; ,
\label{eq:A1}
\end{equation}
where $ S^{(2)}_{\kf} (t) $ denotes the double
convolution in momentum space,
\begin{eqnarray*}
S^{(2)}_{\kf} (t) & = & \frac{1}{N^{2}}
\sum_{\kf_{1}, \kf_{2}}
\left[ \Asf(\kf - \kf_{1}, t)
\, \Asf(\kf_{1} - \kf_{2}, t)
\, \Bsf^{\ast}(\kf_{2}, t) \right. \\
& & \left. \makebox[1.1cm]{ }
+ \Bsf(\kf - \kf_{1}, t)
\, \Bsf(\kf_{1} - \kf_{2}, t)
\, \Asf^{\ast}(\kf_{2}, t) \right] \;\; ,
\end{eqnarray*}
with functions $ \Asf(\kf , t) $ and
$ \Bsf(\kf , t) $ defined as
\begin{eqnarray}
\Asf(\kf , t) & = & e^{-i \omega^{0}_{\kf} t}
\, \fsf(\omega^{0}_{\kf}) \;\; ,
\label{eq:A2} \\
\Bsf(\kf , t) & = & e^{-i \omega^{0}_{\kf} t}
\, \fsf( - \omega^{0}_{\kf}) \;\; .
\label{eq:A3}
\end{eqnarray}
Recalling that the Fourier transform of a
convolution can be expressed as a product of
Fourier transforms of the integrands, we write
$ \Asf(\kf , t) $ and $ \Bsf(\kf , t) $ as
\begin{eqnarray}
\Asf(\kf , t)
& = & {\cal F}^{-1}_{\kf \rightarrow \Rf}
\left[ \asf(\Rf , t) \right] \;\; ,
\label{eq:A4} \\
\Bsf(\kf , t)
& = & {\cal F}^{-1}_{\kf \rightarrow \Rf}
\left[ \bsf(\Rf , t) \right] \;\; ,
\label{eq:A5}
\end{eqnarray}
and disentangle the double convolution
$ S^{(2)}_{\kf} (t) $ as
\[
S^{(2)}_{\kf} (t) =
{\cal F}^{-1}_{\kf \rightarrow \Rf}
\left[ \asf^{2}(\Rf , t)
\, \bsf^{\ast}(\Rf , t)
+ \bsf^{2}(\Rf , t)
\, \asf^{\ast}(\Rf , t) \right] \;\; .
\]
Inserted in expression (\ref{eq:A1}) this gives
\begin{equation}
\Sigma^{(2)}_{r/a}(\Rf, t)
= \mp i \, U^{2} \, \Theta(\pm t)
\, \left[ \asf^{2}(\Rf , t)
\, \bsf^{\ast}(\Rf , t)
+ \bsf^{2}(\Rf , t)
\, \asf^{\ast}(\Rf , t) \right] \;\; .
           \label{eq:sigma2}
\end{equation}
In short, $ {\cal F}_{\omega \rightarrow t} $
yields a convolution in $ \kf $ space, while
$ {\cal F}_{\kf \rightarrow \Rf} $ disentangles
this convolution into a product.

\vspace{0.2cm}
As regards the third-order self-energy, we
proceed in the same way, starting from expression
(\ref{eq:PH}) for
$ \Sigma^{(3)}_{\mathrm{PH}}(\kf,z) $.
As before, the $ \omega \rightarrow t $ Fourier
transform factorizes the $ z $-dependent term in
the denominator, but the $ \kf $ summations still
do not represent a convolution due to the presence
of the unfactorized singular term
$ 1/(\omega^{0}_{\kf_1 - \kf_3}
- \omega^{0}_{\kf_3} - \omega^{0}_{\kf_1 - \kf_2}
+ \omega^{0}_{\kf_2}) $.
To factorize this term as well, we make use of the
identity
\[
\frac{1}{\epsilon} = \frac{1}{2}
\lim_{\delta \to 0^{+}}
\left( \frac{1}{\epsilon + i \delta}
+ \frac{1}{\epsilon - i \delta} \right) \;\; ,
\]
which holds for any $ \epsilon \neq 0 $, and
\[
\frac{1}{\epsilon \pm i \delta} = \frac{1}{i}
\int_{0}^{\pm \infty}\!\! dt
\, e^{i(\epsilon \pm i \delta )t} \;\; ,
\]
which holds for any $ \delta > 0 $, to represent
$ 1/\epsilon $ as
\begin{eqnarray}
\frac{1}{\epsilon} & = & \frac{1}{2i}
\int_{-\infty}^{\infty}\!\! dt
\, {\mathrm s}(t)
\, e^{i \epsilon t} \;\; , \label{eq:A6} \\
{\mathrm s}(t) & \equiv & \mbox{sgn}(t)
\, e^{- 0^{+} |t|} \;\; . \nonumber
\end{eqnarray}
It is clear that (\ref{eq:A6}) factorizes the
above awkward term into a product of four exponentials.
Altogether we obtain
\[
\Sigma^{(3)}_{\mathrm{PH}}(\Rf,t)
= \pm \, \Theta(\pm t) \, U^{3}
\int_{-\infty}^{\infty}\!\! dt' \, {\mathrm s}(t')
{\cal F}_{\kf \rightarrow \Rf}
\left[ S^{(3)}_{\mathrm{PH}}(\kf, t, t')
\right] \;\; ,
\]
where the upper (lower) sign refers to the retarded
(advanced) quantity, respectively, and
$ S^{(3)}_{\mathrm{PH}}(\kf, t, t') $
denotes the triple $ \kf $-space convolution
\begin{eqnarray*}
S^{(3)}_{\mathrm{PH}}(\kf, t, t')
& = & \frac{1}{N^{3}}
\sum_{\kf_{1}, \kf_{2}, \kf_{3}}
\left[ \Asf(\kf - \kf_{1}, t)
\, \Bsf^{\ast}(\kf_{1} - \kf_{2}, t-t')
\, \Asf(\kf_{2}, t-t') \right. \\
& & \left. \makebox[1.5cm]{ }
+ \Bsf(\kf - \kf_{1}, t)
\, \Asf^{\ast}(\kf_{1} - \kf_{2}, t-t')
\, \Bsf(\kf_{2}, t-t') \right] \\
& & \times \left[ \Asf^{\ast}(\kf_{1} - \kf_{3}, t')
\, \Bsf(\kf_{3}, t')
- \Bsf^{\ast}(\kf_{1} - \kf_{3}, t')
\, \Asf(\kf_{3}, t') \right] \;\; .
\end{eqnarray*}
The particle-particle third-order term
$ \Sigma^{(3)}_{\mathrm{PP}}(\kf,z) $ is obtained
from $ - \Sigma^{(3)}_{\mathrm{PH}}(\kf,z) $ by
changing the sign of three $ \omega^{0}_{\kf} $'s
out of five, viz.\ $ \omega^{0}_{\kf - \kf_1} $,
$ \omega^{0}_{\kf_1 - \kf_2} $, and
$ \omega^{0}_{\kf_1 - \kf_3} $.
Since it is obvious from the above definitions
(\ref{eq:A2}) and (\ref{eq:A3}) of $ \Asf(\kf , t) $
and $ \Bsf(\kf , t) $ that
$ \omega^{0}_{\kf} \rightarrow - \omega^{0}_{\kf} $
implies $ \Asf(\kf , t)
\leftrightarrow \Bsf^{\ast}(\kf , t) $, one obtains
$ S^{(3)}_{\mathrm{PP}}(\kf, t, t') $ from
$ - S^{(3)}_{\mathrm{PH}}(\kf, t, t') $ by the
appropriate replacements of those A's and B's that
have $ \kf - \kf_{1} $, $ \kf_{1} - \kf_{2} $, and
$ \kf_{1} - \kf_{3} $ as their
arguments:
\begin{eqnarray*}
S^{(3)}_{\mathrm{PP}}(\kf, t, t')
& = & - \, \frac{1}{N^{3}}
\sum_{\kf_{1}, \kf_{2}, \kf_{3}}
\left[ \Bsf^{\ast}(\kf - \kf_{1}, t)
\, \Asf(\kf_{1} - \kf_{2}, t-t')
\, \Asf(\kf_{2}, t-t') \right. \\
& & \left. \makebox[1.5cm]{ }
+ \Asf^{\ast}(\kf - \kf_{1}, t)
\, \Bsf(\kf_{1} - \kf_{2}, t-t')
\, \Bsf(\kf_{2}, t-t') \right] \\
& & \times \left[ \Bsf(\kf_{1} - \kf_{3}, t')
\, \Bsf(\kf_{3}, t')
- \Asf(\kf_{1} - \kf_{3}, t')
\, \Asf(\kf_{3}, t') \right] \;\; .
\end{eqnarray*}
Expressing $ \Asf(\kf , t) $ and $ \Bsf(\kf , t) $
in terms of their Fourier transforms,
Eqs.\ (\ref{eq:A4}) and (\ref{eq:A5}), as before,
we decouple the above momentum-space convolutions
and get
\begin{eqnarray}
\Sigma^{(3)}_{r/a}(\Rf, t)
& = & \pm \, U^{3} \, \Theta(\pm t)
\, \left\{ \left[ \asf(\Rf , t)
\, {\mathrm w}_{1}(\Rf , t)
- \bsf(\Rf , t)
\, {\mathrm w}_{1}^{\ast}(\Rf , t)
\right] \right. \nonumber \\
& & \makebox[1.9cm]{}
- \left. \left[ \asf^{\ast}(\Rf , t)
\, {\mathrm w}_{2}(\Rf , t)
+ \bsf^{\ast}(\Rf , t)
\, {\mathrm w}_{3}(\Rf , t)
\right] \right\} \;\; ,
               \label{eq:sigma3}
\end{eqnarray}
where the functions $ {\mathrm w}_{i}(\Rf , t) $
are now given in the form of space-time
convolutions
\begin{equation}
{\mathrm w}_{i}(\Rf , t) = \sum_{\bf R'}
\int_{-\infty}^{\infty}\!\! dt'
\, {\mathrm g}_{i}(\Rf - \Rf', t-t')
\, {\mathrm h}_{i}(\Rf' , t') \;\; ,
\label{eq:A7}
\end{equation}
with $ {\mathrm g}_{i}(\Rf , t) $ and
$ {\mathrm h}_{i}(\Rf , t) $ defined as
\begin{eqnarray*}
{\mathrm g}_{1}(\Rf , t)
& = & \asf(\Rf , t) \, \bsf^{\ast}(\Rf , t)
\;\; , \\
{\mathrm g}_{2}(\Rf , t)
& = & \bsf^{2}(\Rf , t) \;\; , \\
{\mathrm g}_{3}(\Rf , t)
& = & \asf^{2}(\Rf , t) \;\; ,\\
{\mathrm h}_{1}(\Rf , t)
& = & {\mathrm s}(t)
\left[ {\mathrm g}_{1}^{\ast}(\Rf , t)
- {\mathrm g}_{1}(\Rf , t) \right] \;\; , \\
{\mathrm h}_{2}(\Rf , t)
& = & {\mathrm s}(t)
\left[ {\mathrm g}_{2}(\Rf , t)
- {\mathrm g}_{3}(\Rf , t) \right] \;\; , \\
{\mathrm h}_{3}(\Rf , t)
& = & {\mathrm h}_{2}(\Rf , t) \;\; .
\end{eqnarray*}
For the 3rd order we thus need two more Fourier
transforms than for the 2nd order, to decouple
these additional space-time convolutions as well:
\begin{equation}
{\mathrm w}_{i}(\Rf , t) =
{\cal F} \left\{ {\cal F}^{-1}
\left[ {\mathrm g}_{i}(\Rf , t) \right]
\cdot {\cal F}^{-1}
\left[ {\mathrm h}_{i}(\Rf , t) \right] \right\}
\;\; .
\label{eq:A8}
\end{equation}

So, in order to obtain the $n$th-order self-energy
$ \Sigma_{\kf}^{(n)}(\omega \pm i 0^{+}) $
via its space-time representation
$ \Sigma^{(n)}_{r/a}(\Rf, t) $ using the
{\sc fft} technique, we first calculate the
quantities
\begin{eqnarray}
\asf(\Rf , t) & = & {\cal F}_{\kf \rightarrow \Rf}
\left[ \Asf(\kf , t) \right] =
\frac{1}{N} \sum_{\kf} e^{i(\kf \cdot \Rf
- \omega^{0}_{\kf} t)}
\, \fsf(\omega^{0}_{\kf}) \;\; , \label{eq:A9} \\
\bsf(\Rf , t) & = & {\cal F}_{\kf \rightarrow \Rf}
\left[ \Bsf(\kf , t) \right] =
\frac{1}{N} \sum_{\kf} e^{i(\kf \cdot \Rf
- \omega^{0}_{\kf} t)}
\, \fsf(- \omega^{0}_{\kf}) \;\; , \label{eq:A10}
\end{eqnarray}
which requires performing the Fourier transform
$ {\cal F}_{\kf \rightarrow \Rf} $. Then we proceed
to evaluate $ \Sigma^{(n)}_{r/a}(\Rf, t) $.
For the 2nd order this amounts just to forming
products of $ \asf(\Rf , t) $'s and $ \bsf(\Rf , t) $'s,
according to expression (\ref{eq:sigma2}).
For the 3rd order we first form functions
$ {\mathrm g}_{i}(\Rf , t) $ and
$ {\mathrm h}_{i}(\Rf , t) $,
again given by products
of $ \asf(\Rf , t) $'s and $ \bsf(\Rf , t) $'s,
then perform a pair of Fourier transforms
$ \left( {\cal F}^{-1} , {\cal F} \right) $,
to calculate functions $ {\mathrm w}_{i}(\Rf , t) $
according to (\ref{eq:A8}), and finally form
$ \Sigma^{(3)}_{r/a}(\Rf, t) $ according to
expression (\ref{eq:sigma3}). Having thus found
the self-energy in space-time representation,
we apply $ {\cal F}^{-1} $ once again to get
\[
\Sigma_{\kf}^{(n)}(\omega \pm i 0^{+})
= {\cal F}^{-1}
\left[ \Sigma^{(n)}_{r/a}(\Rf, t)
\right] \;\; .
\]

The numerical problem involving space-time to
momentum-frequency transformations (and vice versa)
is solved by considering a finite lattice with
periodic boundary conditions and discretizing the
time axis. Here we take $ 256 \times 256 $ lattice
sites and 1024 time points. As regards the
$ {\bf k} \leftrightarrow {\bf R} $ transformation,
the {\sc fft} is an exact procedure, while the
$ t \leftrightarrow \omega $ transformation involves
approximating the continuous Laplace transform with
the corresponding discrete {\sc fft}.
It would be {\em much\/} more time
consuming to evaluate the space-time convolutions
directly.

The 4th-order self-energy $ \Sigma_{\kf}^{(4)}(z) $
comprises twelve topologically nonequivalent diagrams,
nine of which are numerically different. However,
only four out of these nine can be treated entirely
along the same lines as those of the 2nd and 3rd order,
i.e. completely decoupled and evaluated by a series of
Fourier transforms alone. The remaining five do not
yield multiple space-time convolutions, but double
integrals instead, which can be only partly decoupled
by Fourier transforms. The 4th order thus introduces
new numerical difficulties, but tractable ones.
The {\sc fft} can not do the whole job,
but it does all steps but one.

The functions $ \asf(\Rf , t) $ and $ \bsf(\Rf , t) $,
the building blocks of
$ \Sigma^{(n)}_{r/a}(\Rf, t) $, have a
clear physical interpretation as double-time correlation
functions
\begin{eqnarray*}
\asf(\Rf , t)
& = & <\! c_{\bf 0}^{\dagger }(0)
\, c_{\Rf}(t) \! > \;\; , \\
\bsf(\Rf , t)
& = & <\! c_{\Rf}(t)
\, c_{\bf 0}^{\dagger }(0) \! > \;\; ,
\end{eqnarray*}
which can be shown by a straightforward evaluation
of these thermal averages over the eigenstates of
$ H_{0} $, which gives (\ref{eq:A9}) and
(\ref{eq:A10}). This relates them to the unperturbed
Green's functions in space-time representation,
viz.\ the retarded/advanced one,
\begin{eqnarray*}
G^{0}_{r/a}(\Rf , t)
& = & \mp \, i \, \Theta(\pm t)
\, <\! [ c_{\bf 0}^{\dagger }(0),
c_{\Rf}(t) ]_{+} \! > \\
& = & \mp \, i \, \Theta(\pm t)
\, [ \bsf(\Rf , t) + \asf(\Rf , t) ] \;\; ,
\end{eqnarray*}
and the causal one,
\begin{eqnarray*}
G^{0}(\Rf , t) & = & - i \, <\! T\{ c_{\Rf}(t)
\, c_{\bf 0}^{\dagger }(0) \} \! > \\
& = & - i \, [ \Theta(t) \, \bsf(\Rf , t)
- \Theta(-t) \, \asf(\Rf , t) ] \;\; .
\end{eqnarray*}
It is clear that $ \bsf(\Rf , t) $ is the probability
amplitude for an electron to be created at
$ \Rf = {\bf 0} $ and $ t = 0 $, to propagate to the
site \Rf\ which is reached at a later time $ t $, and
to be destructed at $ (\Rf , t) $. This newly created
electron is ``composed of'' the unoccupied (i.e.,
still available) \kf\ states, as indicated by the
amplitudes $ \fsf(- \omega^{0}_{\kf}) =
1 - \fsf(\omega^{0}_{\kf}) $ in the defining relation
(\ref{eq:A10}). In the same way $ \asf(\Rf , t) $
effectively describes the propagation of a hole
created at $ (\Rf , t < 0) $ and destructed at
$ (\Rf = {\bf 0}, t=0) $.

\section*{Acknowledgments}

This work has been financially supported by
the Alexander von Humboldt Foundation,
the Sonderforschungsbereich 166 Duisburg/Bochum,
and the National Science Foundation under grant DMR-9973225.




\newpage

\section*{Figure captions}

\mbox{ } \\
Fig.\ 1. Diagrammatic representation of second- and
third-order self-energy contributions.

\mbox{ } \\
Fig.\ 2.
Real part of
$ \mathrm{\Sigma^{(3)}_{\mathbf{\,\,k}}(\omega)} $
along the
$ \mathrm{\Gamma\rightarrow X}$,
$ \mathrm{X\rightarrow Z} $, and
$ \mathrm{Z\rightarrow \Gamma} $
cuts through the Brillouin zone.
$ \mathrm{Re\,\Sigma^{(3)}_{PH}} $,
$ \mathrm{-Re\,\Sigma^{(3)}_{PP}} $, and
$ \mathrm{Re\,\Sigma^{(3)}
= Re\,\Sigma^{(3)}_{PH}+Re\,\Sigma^{(3)}_{PP}} $
are represented by dashed, dashed-dotted, and
full lines, respectively.

\mbox{ } \\
Fig.\ 3.
Imaginary part of
$ \mathrm{\Sigma^{(3)}_{\mathbf{\,\,k}}(\omega)} $
along the
$ \mathrm{\Gamma\rightarrow X} $,
$ \mathrm{X\rightarrow Z} $, and
$ \mathrm{Z\rightarrow \Gamma} $
cuts through the Brillouin zone.
$ \mathrm{Im\,\Sigma^{(3)}_{PH}} $,
$ \mathrm{- Im\,\Sigma^{(3)}_{PP}} $, and
$ \mathrm{Im\,\Sigma^{(3)}
= Im\,\Sigma^{(3)}_{PH}+Im\,\Sigma^{(3)}_{PP}} $
are represented by dashed, dashed-dotted, and
full lines, respectively.

\mbox{ } \\
Fig.\ 4. Imaginary parts of
$ \mathrm{\Sigma^{(2)}_{\mathbf{\,\,k}}(\omega)} $ and
$ \mathrm{\Sigma^{(3)}_{\mathbf{\,\,k}}(\omega)} $
along the
$ \mathrm{\Gamma\rightarrow X} $,
$ \mathrm{X\rightarrow Z} $,
and $ \mathrm{Z\rightarrow \Gamma} $
cuts through the Brillouin zone.
$ \mathrm{Im\,\Sigma^{(2)}} $,
$ \mathrm{Im\,\Sigma^{(3)}} $, and the sum
$ \mathrm{Im\,\Sigma^{(2)} + Im\,\Sigma^{(3)}} $
are represented by dashed, dashed-dotted, and full
lines, respectively.

\mbox{ } \\
Fig.\ 5. Imaginary parts of
$ \mathrm{\Sigma^{(2)}_{\mathbf{\,\,k}}(\omega)} $ and
$ \mathrm{\Sigma^{(3)}_{\mathbf{\,\,k}}(\omega)} $ at
$ \mathrm{\Gamma} $, X, Z, and M points close to
$ \omega = E_{F} $.
$ \mathrm{Im\,\Sigma^{(2)}} $,
$ \mathrm{Im\,\Sigma^{(3)}} $, and the sum
$ \mathrm{Im\,\Sigma^{(2)} + Im\,\Sigma^{(3)}} $
are represented by dashed, dashed-dotted, and full
lines, respectively.

\mbox{ } \\
Fig.\ 6. Single-particle spectral function
$ \mathrm{A_{\mathbf{k}}(\omega)} $ along the
$ \mathrm{\Gamma\rightarrow X} $,
$ \mathrm{X\rightarrow Z} $, and
$ \mathrm{Z\rightarrow \Gamma} $ cuts through the
Brillouin zone.

\mbox{ } \\
Fig.\ 7. Quasiparticle dispersion (circles) derived
from the maxima of the spectral function
$ \mathrm{A_{\mathbf{k}}(\omega)} $ and the
unrenormalized dispersion (full line).

\mbox{ } \\
Fig.\ 8. Renormalized (full line) and unrenormalized
(dashed line) density of states.

\newpage

\begin{figure}

 \begin{center}
  \begin{minipage}{17cm}

   \vspace{1cm}
   \begin{minipage}[h]{17cm}
    \begin{center}
     \epsfxsize=8cm
     \epsfbox{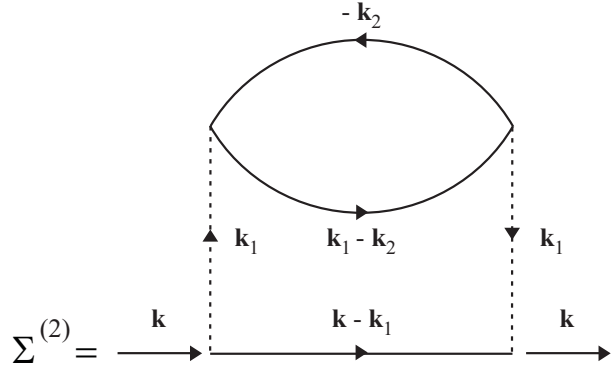}
    \end{center}
   \end{minipage}

   \vspace{1cm}
   \begin{minipage}[h]{17cm}
    \begin{center}
     \epsfxsize=8cm
     \epsfbox{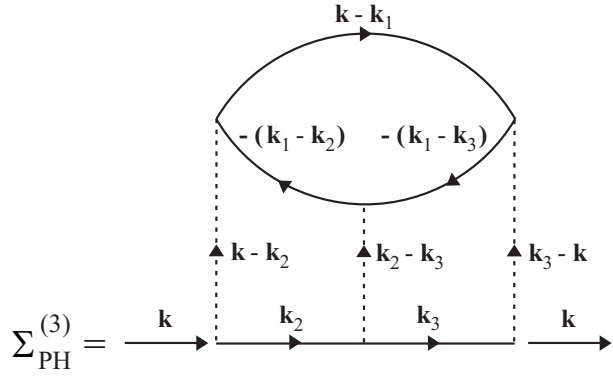}
    \end{center}
   \end{minipage}

   \vspace{1cm}
   \begin{minipage}[h]{17cm}
    \begin{center}
     \epsfxsize=8cm
     \epsfbox{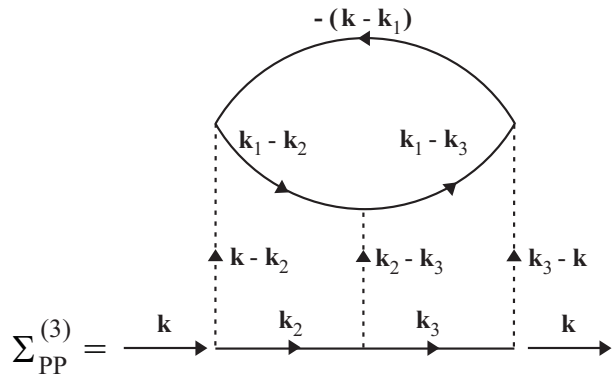}
    \end{center}
   \end{minipage}

  \end{minipage}
 \end{center}

 \caption{Diagrammatic representation of second- and
third-order self-energy contributions.}
\end{figure}

\newpage

\begin{figure}

 \begin{center}
  \begin{minipage}{19.5cm}

   \vspace{-1.75cm}
   \hspace{-1.5cm}
    \begin{minipage}[t]{17.5cm}
     \begin{center}
       \begin{minipage}{17.5cm}
         \epsfxsize=17.5cm
         \epsfbox{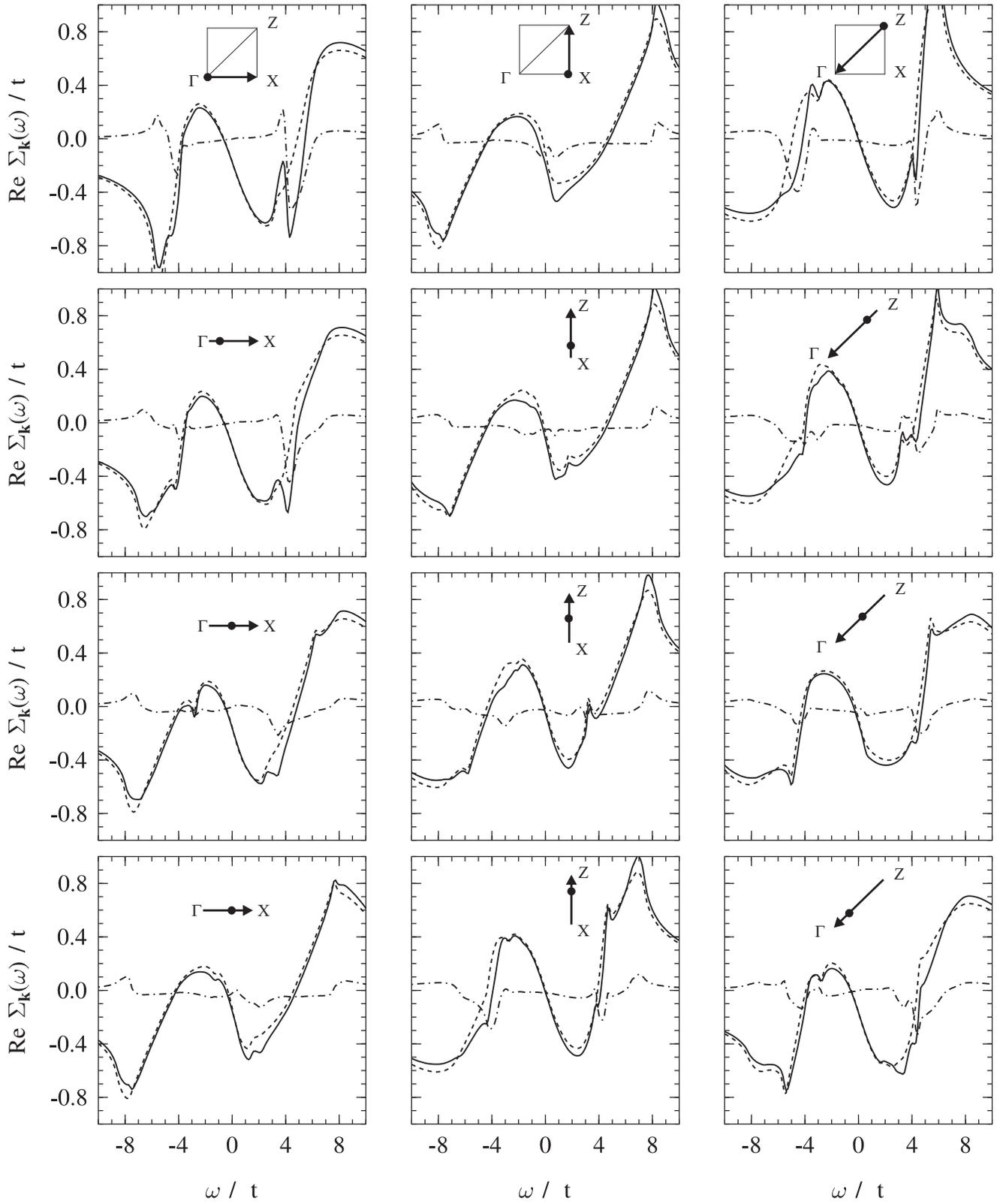}
        \end{minipage}
     \end{center}
   \end{minipage}

  \end{minipage}
 \end{center}

 \caption{Real part of
$ \mathrm{\Sigma^{(3)}_{\mathbf{\,\,k}}(\omega)} $
along the
$ \mathrm{\Gamma\rightarrow X}$,
$ \mathrm{X\rightarrow Z} $, and
$ \mathrm{Z\rightarrow \Gamma} $
cuts through the Brillouin zone.
$ \mathrm{Re\,\Sigma^{(3)}_{PH}} $,
$ \mathrm{-Re\,\Sigma^{(3)}_{PP}} $, and
$ \mathrm{Re\,\Sigma^{(3)}
= Re\,\Sigma^{(3)}_{PH}+Re\,\Sigma^{(3)}_{PP}} $
are represented by dashed, dashed-dotted, and
full lines, respectively.}

\end{figure}

\newpage

\begin{figure}

 \begin{center}
  \begin{minipage}{19.5cm}

   \vspace{-1.75cm}
   \hspace{-1.5cm}
    \begin{minipage}[t]{17.5cm}
     \begin{center}
       \begin{minipage}{17.5cm}
         \epsfxsize=17.5cm
         \epsfbox{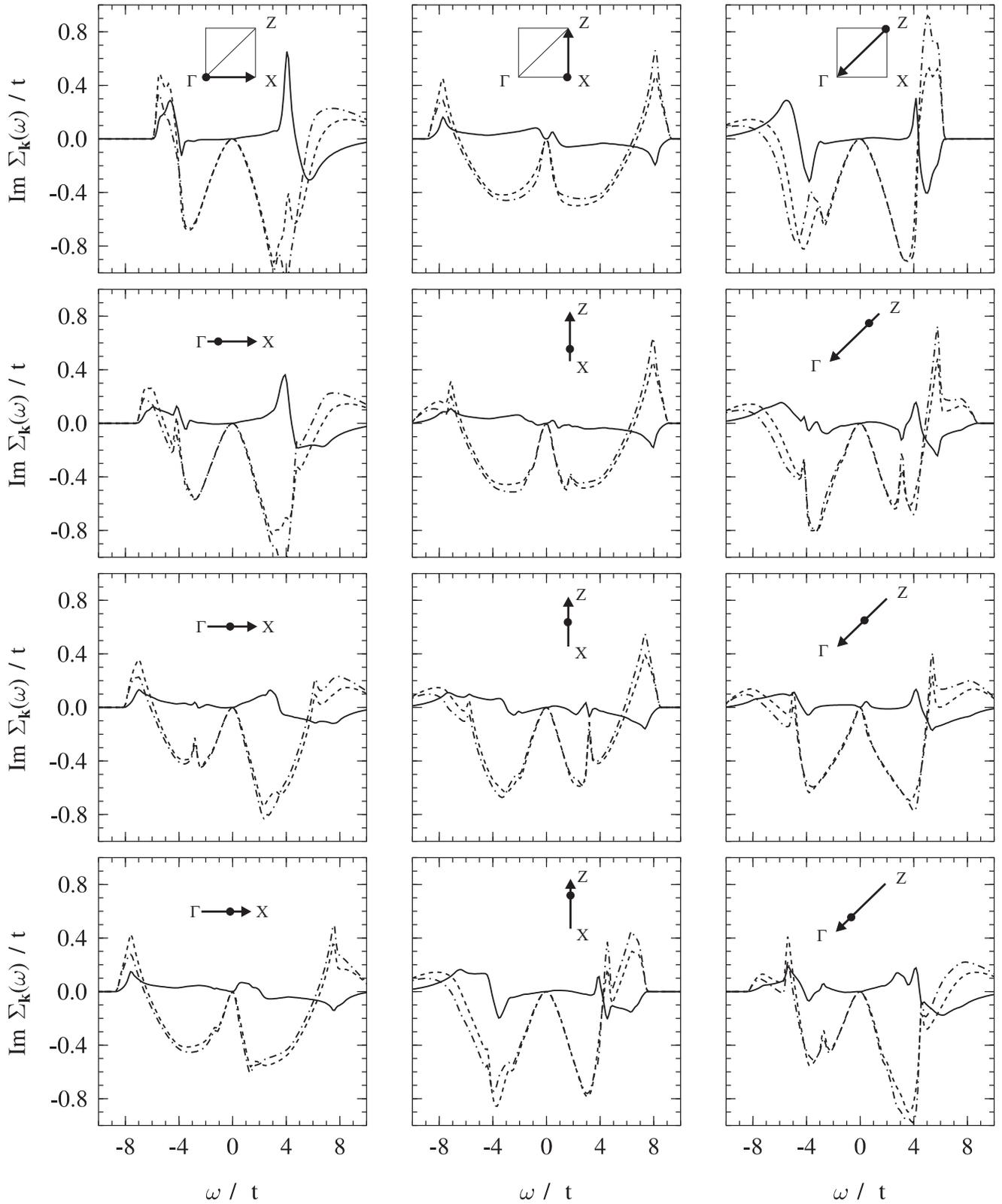}
        \end{minipage}
     \end{center}
   \end{minipage}

  \end{minipage}
 \end{center}

 \caption{Imaginary part of
$ \mathrm{\Sigma^{(3)}_{\mathbf{\,\,k}}(\omega)} $
along the
$ \mathrm{\Gamma\rightarrow X} $,
$ \mathrm{X\rightarrow Z} $, and
$ \mathrm{Z\rightarrow \Gamma} $
cuts through the Brillouin zone.
$ \mathrm{Im\,\Sigma^{(3)}_{PH}} $,
$ \mathrm{- Im\,\Sigma^{(3)}_{PP}} $, and
$ \mathrm{Im\,\Sigma^{(3)}
= Im\,\Sigma^{(3)}_{PH}+Im\,\Sigma^{(3)}_{PP}} $
are represented by dashed, dashed-dotted, and
full lines, respectively.}

\end{figure}

\newpage

\begin{figure}

 \begin{center}
  \begin{minipage}{19.5cm}

   \vspace{-1.75cm}
   \hspace{-1.5cm}
    \begin{minipage}[t]{17.5cm}
     \begin{center}
       \begin{minipage}{17.5cm}
         \epsfxsize=17.5cm
         \epsfbox{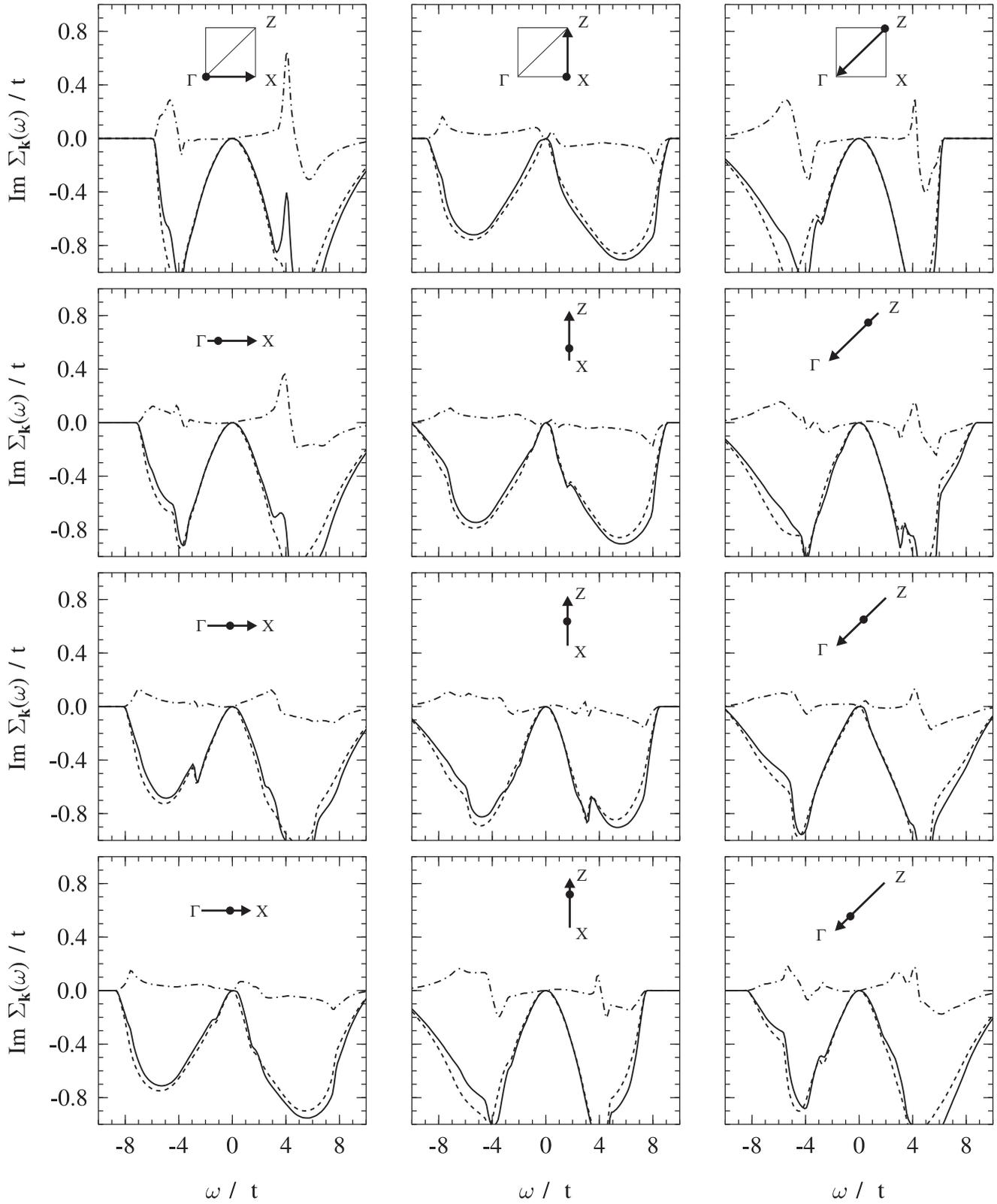}
        \end{minipage}
     \end{center}
   \end{minipage}

  \end{minipage}
 \end{center}

 \caption{Imaginary parts of
$ \mathrm{\Sigma^{(2)}_{\mathbf{\,\,k}}(\omega)} $ and
$ \mathrm{\Sigma^{(3)}_{\mathbf{\,\,k}}(\omega)} $
along the
$ \mathrm{\Gamma\rightarrow X} $,
$ \mathrm{X\rightarrow Z} $,
and $ \mathrm{Z\rightarrow \Gamma} $
cuts through the Brillouin zone.
$ \mathrm{Im\,\Sigma^{(2)}} $,
$ \mathrm{Im\,\Sigma^{(3)}} $, and the sum
$ \mathrm{Im\,\Sigma^{(2)} + Im\,\Sigma^{(3)}} $
are represented by dashed, dashed-dotted, and full
lines, respectively.}

\end{figure}

\newpage

\begin{figure}

 \begin{minipage}{17.5cm}
  \begin{center}
   \epsfxsize=6cm
   \epsfbox{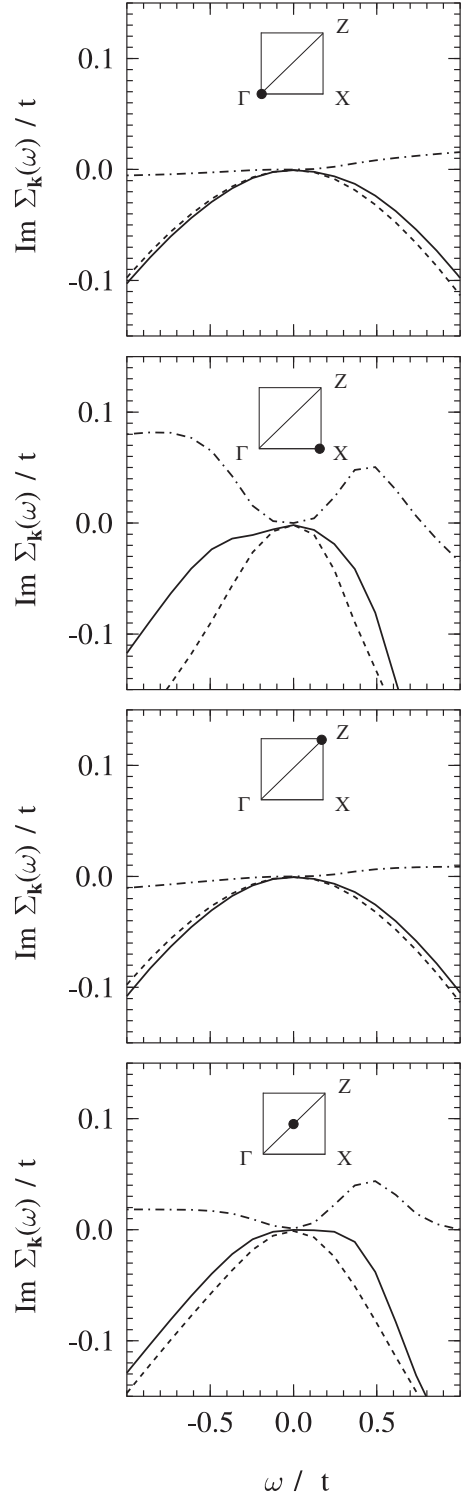}
  \end{center}
 \end{minipage}

\caption{Imaginary parts of
$ \mathrm{\Sigma^{(2)}_{\mathbf{\,\,k}}(\omega)} $ and
$ \mathrm{\Sigma^{(3)}_{\mathbf{\,\,k}}(\omega)} $ at
$ \mathrm{\Gamma} $, X, Z, and M points close to
$ \omega = E_{F} $.
$ \mathrm{Im\,\Sigma^{(2)}} $,
$ \mathrm{Im\,\Sigma^{(3)}} $, and the sum
$ \mathrm{Im\,\Sigma^{(2)} + Im\,\Sigma^{(3)}} $
are represented by dashed, dashed-dotted, and full
lines, respectively.}

\end{figure}

\newpage

\begin{figure}

 \begin{center}
  \begin{minipage}{19.5cm}

   \vspace{-1.75cm}
   \hspace{-1.5cm}
    \begin{minipage}[t]{17.5cm}
     \begin{center}
       \begin{minipage}{17.5cm}
         \epsfxsize=17.5cm
         \epsfbox{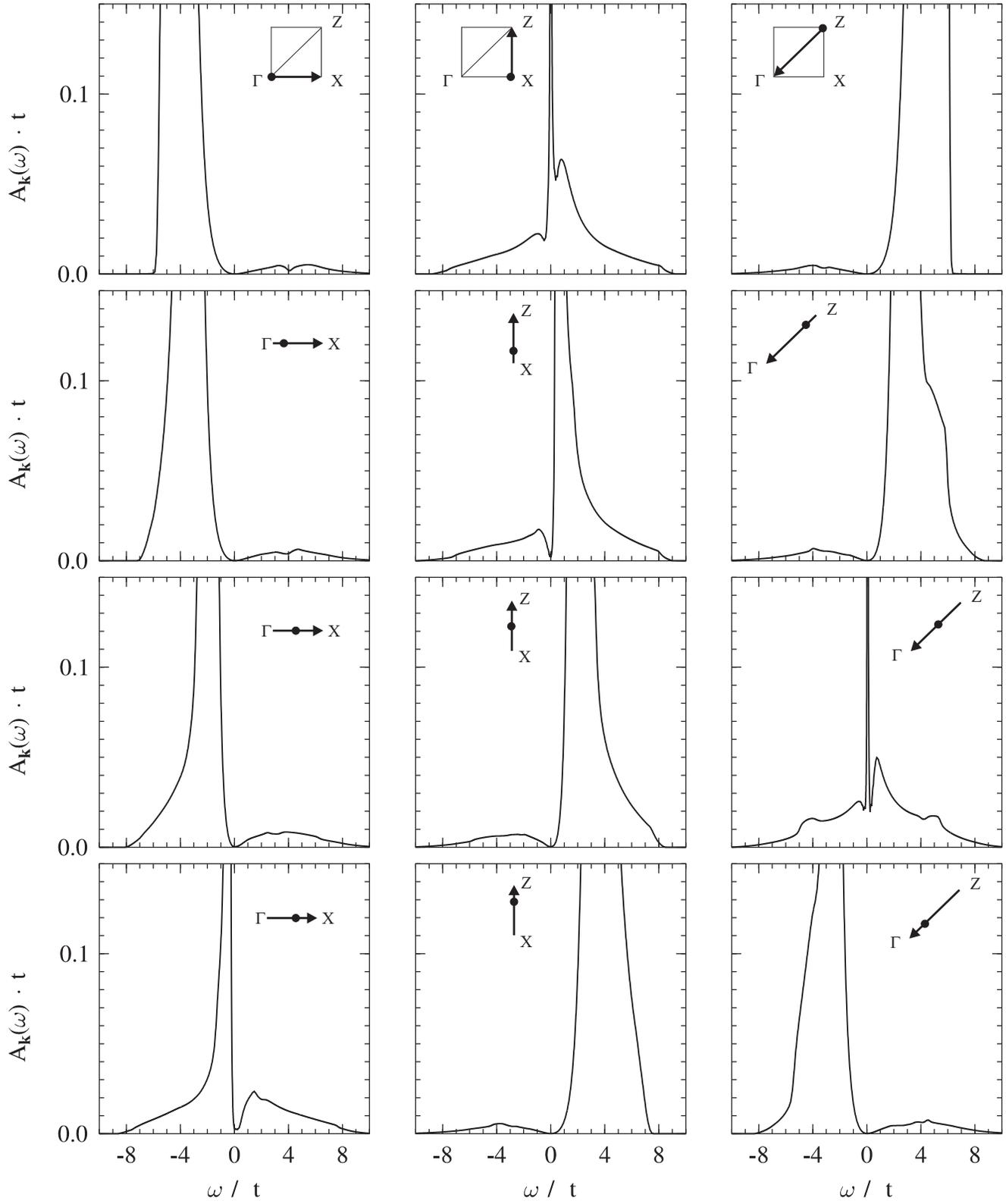}
        \end{minipage}
     \end{center}
   \end{minipage}

  \end{minipage}
 \end{center}

 \caption{Single-particle spectral function
$ \mathrm{A_{\mathbf{k}}(\omega)} $ along the
$ \mathrm{\Gamma\rightarrow X} $,
$ \mathrm{X\rightarrow Z} $, and
$ \mathrm{Z\rightarrow \Gamma} $ cuts through the
Brillouin zone.}

\end{figure}

\newpage

\begin{figure}

 \begin{minipage}[t]{17.5cm}
  \begin{center}
   \epsfxsize=10cm
   \epsfbox{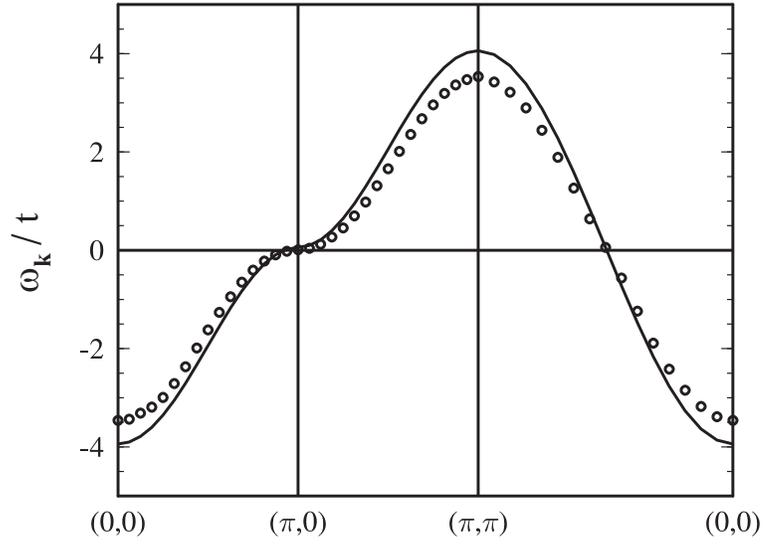}
  \end{center}
 \end{minipage}

 \caption{Quasiparticle dispersion (circles) derived
from the maxima of the spectral function
$ \mathrm{A_{\mathbf{k}}(\omega)} $ and the
unrenormalized dispersion (full line).}

\end{figure}

\newpage

\begin{figure}

 \begin{minipage}{17.5cm}
  \begin{center}
   \epsfxsize=10cm
   \epsfbox{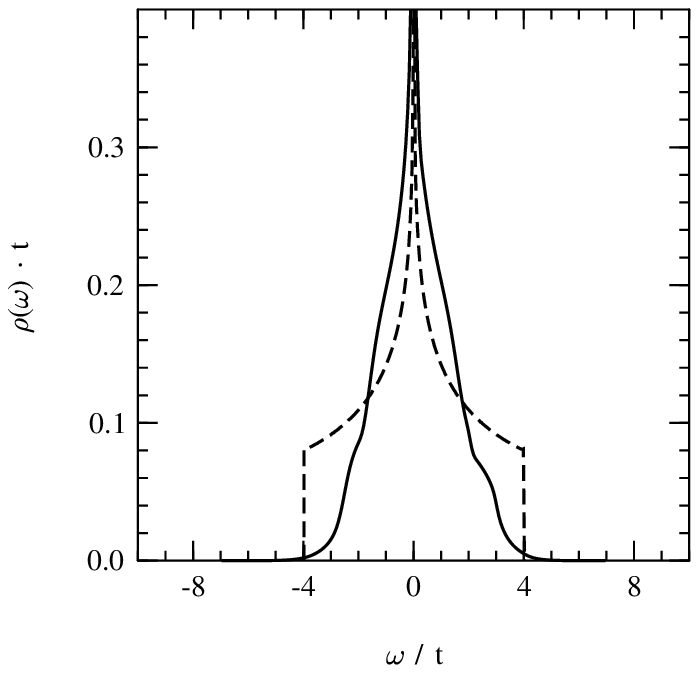}
  \end{center}
 \end{minipage}

 \caption{Renormalized (full line) and unrenormalized
(dashed line) density of states.}

\end{figure}


\begin{thebibliography}{99}
\bibitem{comment} This could be due to three
unfavourable possibilities:
(i) the expansion might be an asymptotic one, and
therefore irrelevant for any value of $ U $ except
$ U \rightarrow 0 $;
(ii) the expansion might be convergent, but the
radius of convergence might be exceeded;
(iii) even within the radius of convergence the
series might converge very slowly.
\bibitem{error}
The small shift seen in the numerical data for
some {\bf k} points is caused by the damping factor
in the unperturbed propagators, which is introduced
to speed up the calculations.
\end{thebibliography}
\end{document}